\documentclass[final,5p,twocolumn]{elsarticle}
\usepackage{graphics}
\usepackage{graphicx}
\usepackage{epsfig}
\usepackage{amssymb}
\usepackage{amsmath}
\usepackage{hyperref}

\def\fmn#1#2{\mbox{${\textstyle \frac{#1}{#2}}$}}
\journal{Physics Letters B}

\begin{document}
\begin{frontmatter}
\title{The analysing powers in proton-deuteron elastic scattering}

\author[dubna,moscow,dubnau]{Yu.~Uzikov}\ead{uzikov@jinr.ru}
\author[london]{C.~Wilkin}\ead{c.wilkin@ucl.ac.uk}

\address[dubna]{Laboratory of Nuclear Problems, JINR, RU-141980 Dubna, Russia}
\address[moscow]{Department of Physics, M.~V.~Lomonosov Moscow State University, RU-119991 Moscow, Russia}
\address[dubnau]{Dubna State University, RU-141980 Dubna, Russia}
\address[london]{Physics and Astronomy Department, UCL, Gower Street, London WC1E 6BT, United Kingdom}

\begin{abstract}
It is shown that the ratio of the deuteron and proton analysing powers in
proton-deuteron elastic scattering at small angles is sensitive to subtle
effects in a theoretical description. These include the transverse
spin-spin term in the elementary nucleon-nucleon amplitudes and
double-scattering corrections. On the other hand there is far less
sensitivity to the spin-orbit amplitude and to binding or other kinematic
effects associated with the use of the deuteron, as either target or
projectile. The available data are in agreement with the results of a
refined Glauber theory model.
\end{abstract}
\end{frontmatter}

In the analysis of their data on proton-deuteron elastic scattering at
796~MeV, the authors of Ref.~\cite{IRO1983} pointed out that, at small c.m.\
momentum transfer $\vec{q}$ between the initial and final proton momenta, the
proton analysing power $A_y^p$ is dominantly determined by an interference of
charge-average spin-independent mucleon-nucleon amplitudes with the
corresponding spin-orbit term. This is in contrast to the differential cross
section, which is significantly reduced by the interaction of the proton with
both constituents of the deuteron~\cite{IRO1983}. If this approach provides a
good approximation for $A_y^p$, one should check whether it leads to a
reasonable description of the deuteron vector analysing power $A_y^d$. It is
the purpose of this note to compare the values of $A_y^p$ and $A_y^d$ in $pd$
elastic scattering by evaluating the ratio
\begin{equation}
\label{ratio}
R = A_y^d/A_y^p
\end{equation}
at various beam energies and momentum transfers in order to investigate
deviations from the simple model proposed in~\cite{IRO1983} which, as
discussed in the Appendix, would suggest that $R=2/3$.

\begin{figure}[htb]
\begin{center}
\includegraphics[width=1.0\columnwidth]{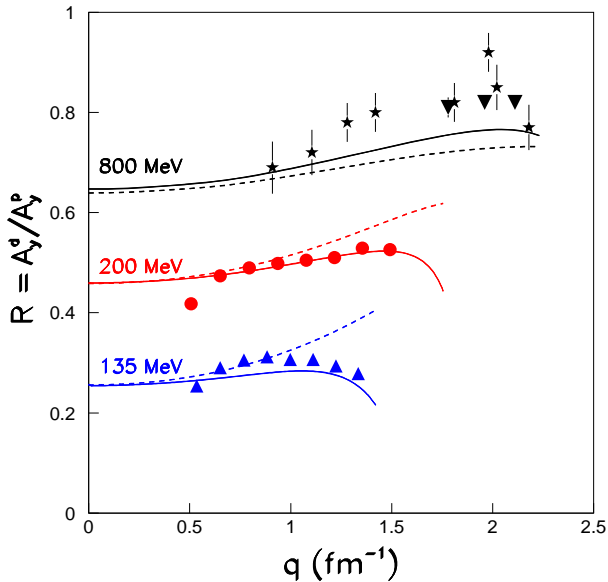}
\caption{\label{figure} Values of the ratio $R$ of the deuteron to  proton
analysing powers measured in proton-deuteron elastic scattering. In order to
minimise the confusion caused by the overlap of data obtained at the
different proton beam energies noted, the data and calculation at 200 and
135~MeV have been lowered by 0.2 and 0.4, respectively. The (black) star
values at $T_p=800$~MeV were obtained by combining the $A_y^d$ data of
Ref.~\cite{HAJ1987} with the $A_y^p$ data of Ref.~\cite{BAR2018}. The
(black) inverted triangles used instead the later $A_y^d$ data of the same
group~\cite{GHA1991}. If the results of Ref.~\cite{IRO1983} had been used for
$A_y^p$ the experimental values of $R$ would be lowered by about 3\%.
The data at 135~MeV (blue triangles) and 200~MeV (red circles) are IUCF
measurements~\cite{PRZ2006}, where both the deuteron and proton analysing
powers were studied in the same experiment. The curves were evaluated in the
refined multiple scattering scheme~\cite{PLA2010} using all the spin dependence
of the $NN$ amplitudes as determined in a partial wave analysis~\cite{ARN2007}.
The dashed lines correspond to the single-scattering approximation, whereas the
solid lines represent the full model of Ref.~\cite{PLA2010}. In general the full
model describes the $q$ dependence quite well, though the first IUCF
points~\cite{PRZ2006} are more problematic, especially given that $R$ must vary
like $A+Bq^2$ for small values of $q$,
}
\end{center}
\end{figure}

Any investigation of $R$ is hampered by a lack of data on either $A_y^p$ or
$A_y^d$ at similar energies per nucleon. However at 796~MeV, in addition to
the $A_y^p$ data given in~\cite{IRO1983}, there are also measurements from
COSY-ANKE~\cite{BAR2018}. These are complemented by measurements of the
deuteron analysing power in $dp$ elastic scattering in the 800~MeV per
nucleon region~\cite{HAJ1987,ARV1988,GHA1991}. The resulting values of $R$
are reported in Fig.~\ref{figure} as function of the magnitude of the
momentum transfer $q$. The error bars are statistical and do not take into
account the systematic uncertainties associated with the beam polarisations
in the various experiments.

By using a polarised deuterium target, together with a polarised proton beam,
it was possible at IUCF to measure both the proton and deuteron analysing
powers in $pd$ elastic scattering in the same experiment~\cite{PRZ2006}. The
results obtained at 135 and 200~MeV are also shown in Fig.~\ref{figure}
though, for clarity of presentation, these have been displaced downwards by
$0.4$ and $0.2$, respectively.

The only other published data where the ratio can be evaluated were taken at
250~MeV per nucleon~\cite{HAT2002,SEK2011} but, due to the lack of small
angle deuteron data, only two points could be used and these yielded
$R=0.65\pm0.03$ and $R=0.66\pm0.01$ at $q=1.45$~fm$^{-1}$ and
$1.59$~fm$^{-1}$, respectively. These values are clearly compatible with $R=2/3$,
especially if one adds to the statistical errors the 3\% systematic uncertainty
in the beam polarisations.

The data shown in Fig.~\ref{figure} are more or less consistent with the
simple-minded expectation of $R=2/3$ at low $q$ but the values of $R$ appear
not to be constant. It is therefore of interest to see what $q$ dependence is
to be expected in more realistic theoretical models. The most transparent
approach, especially at the higher energies, is a generalisation of the
Glauber eikonal model~\cite{GLA1955}. Here all the spin-dependence of the
nucleon-nucleon amplitudes is retained~\cite{PLA2010}, though the
cancellation of the higher order terms~\cite{HAR1969} can no longer be
guaranteed because of the non-commutativity of some of the amplitudes. Though
this approach has been used to describe the differential cross
section~\cite{FRI2018}, it can also be used in the study of polarisation
observables~\cite{PLA2010}. For example, the individual analysing powers have
already been studied in this model at 135~MeV and 200~MeV~\cite{TEM2015}.

In the Glauber model, the $pd\to pd$ elastic scattering amplitude contains
terms corresponding to single and double scattering of the proton on,
respectively, one or two nucleons in the deuteron. The original
approach~\cite{GLA1955} neglected spin degrees of freedom of the particle
scattering from the deuteron. These have been included by later authors in a
refined Glauber model~\cite{PLA2010,ALB1982}, where the single and double
scattering are evaluated using the full spin dependence of the on-shell
nucleon-nucleon amplitudes. Because of the size of the deuteron, single
scattering dominates at small momentum transfer $\vec{q}$ but the size also
means that the single scattering falls fast with $q$ and the double
scattering, where the momentum transfer is shared between the two nucleons in
the deuteron, then plays a more important role.

In the single-scattering approximation the proton or deuteron analysing
powers at small $q$ result from an interference between the $NN$
charge-average spin-orbit amplitude and combinations of the spin-independent
and transverse spin-spin amplitudes~\cite{PLA2010}. The value of $R=2/3$ at
small $q$ follows after neglecting the spin-spin term\footnote{The 2/3 factor
is discussed further in the Appendix.}. Given that this amplitude is
generally small~\cite{ARN2007}, it is not surprising that the
single-scattering predictions in Fig.~\ref{figure} are close to $2/3$. The
deviations from this value are generally much better reproduced by the full
(single + double-scattering) model than the single-scattering calcullation,
though at 800~MeV there seems to be a systematic difference of about 7\%.
This may be largely due to uncertainties in the beam polarisation and,
indeed, if the earlier $A_y^p$ values~\cite{IRO1983} had been used, the
discrepancy would be reduced to about 4\%.

As seen in Fig.~\ref{figure}, even in the single scattering approximation
there is a $q$ dependence arising, among other things, from the small $NN$
amplitudes and the deuteron $D$-state. Although both $A_y^p$ and $A_y^d$
vanish as $q\to 0$, their ratio approaches a finite limit. In
Table~\ref{forward} are shown the predictions for $R$ at $q=0$ in the refined
Glauber model~\cite{PLA2010} for a variety of proton beam energies where
there are either proton or deuteron data. The table shows that, at low values
of $q$, there is very little difference between the predictions of the full
model and those where only single scattering is considered.

\begin{table}[h!]
\caption{Predicted values of the ratio of deuteron to proton analysing powers
in $pd$ elastic scattering as $q\to 0$. The single ($SS$) and full ($SS+DS$)
models of Ref.~\cite{PLA2010} were evaluated using as input a partial wave
analysis of the nucleon-nucleon amplitudes~\cite{ARN2007}. The table shows
the small deviations of $R$ from $2/3$.
\label{forward} \vspace{3mm} }%
\centering
\begin{tabular}{|r|r|r|}
\hline
$T_p\phantom{a}$&\multicolumn{2}{|c|}{$100(R-2/3)$}\\
\cline{2-3}
MeV&SS\phantom{ll}&SS+DS\\
\hline
135&$-1.09$&$-1.24\phantom{0}$\\
200&$-0.82$&$-0.73\phantom{0}$\\
250&$-1.02$&$-0.81\phantom{0}$\\
450&$-2.25$&$-1.55\phantom{0}$\\
600&$-4.28$&$-3.31\phantom{0}$\\
800&$-2.75$&$-2.00\phantom{0}$\\
1000&$-0.36$&$0.25\phantom{0}$\\
1125&$1.84$&$2.35\phantom{0}$\\
1135&$2.04$&$2.53\phantom{0}$\\
\hline
\end{tabular}
\end{table}

It follows from the results given in Table~\ref{forward} that, within the
refined Glauber model, most of the deviations of $R$ from 2/3 at $q=0$ are
due to the spin-spin term in single scattering; the modifications due to the
double scattering are small in comparison and may be estimated from theory
with sufficient precision. Using the notation of Platonova and
Kukulin~\cite{PLA2010}, the single-scattering approximation gives, to lowest
order in $A_{10}/A_{1}$,
\begin{equation}
\label{Yuri}
Re\{A_2^*A_{10}\}/Re\{A_2^*A_1\} = \fmn{9}{2}(R-\fmn{2}{3}).
\end{equation}
Here $A_1$ is the dominant spin-independent amplitude, $A_2$ is the
spin-orbit amplitude, and $A_{10}$ is the transverse spin-spin amplitude, all
in the limit of $q\to 0$. The small relativistic correction to the spin-orbit
amplitude~\cite{SOR1979,PLA2010} vanishes as $q\to 0$ and has here been
neglected.

It can be seen from Eq.~(\ref{Yuri}) that the ratio is independent of the
size of the spin-orbit amplitude and so binding corrections to the spin-orbit
amplitude are of little importance. As a consequence, a precise measurement
of $R$ could provide some information on the $NN$ transverse spin-spin
amplitude in the forward direction that is independent of the extraction of
the imaginary part of amplitude via the spin dependence of total cross
sections and the evaluation of the corresponding real part from forward
dispersion relations~\cite{GRE1982}.

The analysing powers of both the proton and deuteron in the $pd\to pd$
scattering have similar shapes as functions of $q$~\cite{PRZ2006}. However,
due in part to the multiple scattering, the position of the zeroes in $A_y^p$
and $A_y^d$ are displaced slightly and so in this region their ratio can
fluctuate strongly. For smaller values of $q$, which are those shown in
Fig.~\ref{figure}, the ratio $R$ varies little from its standard value of
2/3. Nevertheless, its $q$ dependence is quite similar to that expected on
the basis of the refined Glauber model, where the spin dependence of the
double scattering is taken seriously~\cite{PLA2010}.

\begin{figure}[htb]
\begin{center}
\includegraphics[width=1.0\columnwidth]{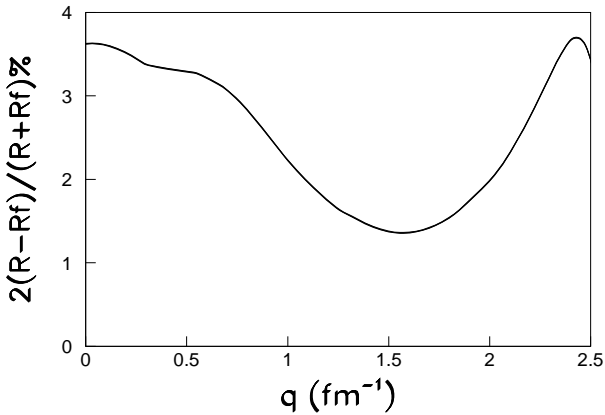}
\caption{\label{figure2}Difference between the predictions of the refined
Glauber model~\cite{PLA2010} without ($R$) and with ($Rf$) the $NN$
spin-spin contribution at 800~MeV expressed as a percentage of their average.
}
\end{center}
\end{figure}

At 800~MeV the predictions of the ratio $R$ in the refined Glauber model are
better than those for the individual proton and deuteron analysing
powers~\cite{PLA2018}. This is to be contrasted with the good agreement found
for both analysing powers at both 135~MeV and 200~MeV~\cite{TEM2015}. This
brings into question the reliability of the input $NN$ amplitudes at the
higher energy, If the proton and deuteron polarisations can be well
controlled at the 1\% level then the value of $R$ could yield information on
the nucleon-nucleon amplitudes. This is illustrated in Fig.~\ref{figure2},
which shows the predicted influence of the nucleon-nucleon spin-spin
amplitudes on the values of $R$. Typically the effects are on the 2-3\% level
and measurements must be better than this to yields a meaningful constraint
on the spin-spin amplitudes.

%
%
\section*{{Acknowledgments}} Correspondence with E.~Bleszynski,
M.~Bleszynski, V.~Ghazikhanian, and M.~Platonova has been very helpful.
%
%

\section*{Appendix: The single-scattering model}

The simplest model that can generate both proton and deuteron non-zero
analysing powers in $dp$ elastic scattering has a transition operator of the
form
\begin{equation}
\label{a1}
 \hat M=a+ ib \hat{\sigma}_y + ic \hat{S}_y.
\end{equation}
Here $\hat \sigma_y$ and  $\hat S_y$ are operators acting, respectively, on
the spins of the proton and deuteron. The proton analysing power results from
an interference between the amplitudes $a$ and $b$ whereas that of the
deuteron is due to an interference between $a$ and $c$. Straightforward
calculations yield
\begin{eqnarray}
\nonumber
A_y^p &=& 2 Im\{ab^*\}/[|a|^2+|b|^2 +\fmn{2}{3}|c|^2],\\
A_y^d &=& \fmn{4}{3} Im\{ac^*\}/[|a|^2+|b|^2 +\fmn{2}{3}|c|^2].
\end{eqnarray}
However, in the single scattering approximation at low $q$, neglecting the
small relativistic correction~\cite{SOR1979}, the spin-dependent amplitudes
are equal, $b=c$~\cite{PLA2010}. It then follows that $A_y^d =
\fmn{2}{3}A_y^p$, where the $2/3$ factor actually arises from the fact that
for $A_y^d$ one has to sum over the two spin projections of the proton,
whereas for $A_y^p$ the sum is over the three spin projections of the
deuteron. Deviations from 2/3 at low $q$ are primarily due to the spin-spin
amplitudes and the double scattering, both of which are included in the
refined Glauber model~\cite{PLA2010}.
%
%


\section*{\small{References}}
\begin{thebibliography}{99}
%
\bibitem{IRO1983} F.~Irom, G.J.~Igo, J B.~McClelland, C.A.~Whitten, Jr., M.~Bleszynski, Phys.\ Rev.\ C \ {28} (1983) 2380.
%
\bibitem{BAR2018} S.~Barsov \emph{et al.}, Eur.\ Phys.\ J.\ A 54 (2018) 225.
%
\bibitem{HAJ1987} M.~Haji-Saied \emph{et al.}, Phys.\ Rev.\ C {36} (1987) 2010.
%
\bibitem{ARV1988} J.~Arvieux \emph{et al.}, Nucl.\ Instrum.\ Methods Phys.\ Res.\ A {273} (1988) 48.
%
\bibitem{GHA1991} V.~Ghazikhanian \emph{et al.}, Phys.\ Rev.\ C {43} (1991) 1532.
%
\bibitem{PRZ2006} B.~v.~Przewoski \emph{et al.}, Phys.\ Rev.\ C 74 (2006) 064003.
%
\bibitem{HAT2002} K.~Hatanaka \emph{et al.}, Phys.\ Rev.\ C 66 (2002) 044002.
%
\bibitem{SEK2011} K.~Sekiguchi \emph{et al.}, Phys.\ Rev.\ C 83 (2011) 061001 (R).
%
\bibitem{GLA1955} R.J.~Glauber, Phys.\ Rev.\ 100 (1955) 242.
%
\bibitem{PLA2010} M.N.~Platonova, V.I.~Kukulin, Phys.\ Rev.\ C {81} (2010) 014004.
%
\bibitem{HAR1969} D.R.~Harrington, Phys.\ Rev.\ 184 (1969) 1745.
%
\bibitem{FRI2018} C.~Fritzch \emph{et al.}, Phys.\ Lett.\ B 784 (2018) 277.
%
\bibitem{TEM2015} A.A.~Temerbayev and Yu.N.~Uzikov, Yad.\ Fiz.\ 78 (2015) 38: [Phys.\ Atom.\ Nucl.\ 78 (2015) 35].
%
\bibitem{ARN2007} R.A.~Arndt, W.J.~Briscoe, I.I.~Strakovsky, R.L.~Workman, Phys.\ Rev.\ C 76 (2007) 025209;
{http://gwdac.phys.gwu.edu}.
%
\bibitem{ALB1982} G.~Alberi, M.~Bleszynski, T.~Jaroszewicz, Ann.\ Phys.\ (NY) 142 (1982) 299.
%
\bibitem{SOR1979} C.~Sorensen, Phys.\ Rev.\ D 19 (1979) 1444.
%
\bibitem{GRE1982} W.~Grein, P.~Kroll, Nucl.\ Phys.\ A 377 (1982) 505.
%
\bibitem{PLA2018} M.N.~Platonova, V.I.~Kukulin, submitted to Phys.\ Rev.\ C.
%
\end{thebibliography}
\end{document}